\documentclass[amsmath,amssymb,nofootinbib,prd]{revtex4}

\newcommand{\be}{\begin{equation}}
\newcommand{\ee}{\end{equation}}
\newcommand{\bea}{\begin{eqnarray}}
\newcommand{\eea}{\end{eqnarray}}

\begin{document}

\title{\bf Regular spherical dust spacetimes}

\author{Neil P. Humphreys$^1$, Roy Maartens$^{1,2}$ and David R. Matravers$^1$\\~}

\affiliation{$^1$Institute of Cosmology \& Gravitation, University of Portsmouth, Portsmouth PO1 3FX, UK 
\\ $^2$Physics Department, University of the Western Cape, Cape Town 7535, South Africa}


\begin{abstract}

Physical (and weak) regularity conditions are used to determine
and classify all the
possible types of spherically symmetric
dust spacetimes in general relativity.
This work unifies and
completes various earlier results.
The junction conditions are described
for general non-comoving (and non-null) surfaces,
and the limits of kinematical quantities are given on all comoving
surfaces where there is Darmois matching.
We show that an inhomogeneous generalisation of the
Kantowski-Sachs metric may be joined to the
Lema\^{\i}tre-Tolman-Bondi metric.
All the possible
spacetimes are explicitly divided into four groups according
to topology, including a group in which
the spatial sections have the topology of a 3-torus.
The recollapse conjecture (for these spacetimes)
follows naturally in this approach.  

\end{abstract}

\maketitle

\section*{I. Introduction}

The convenient Lema\^{\i}tre-Tolman-Bondi (LTB) exact solutions
have been exploited as the main inhomogeneous models in relativity
and cosmology for many years. The remarkably rich structure of these
solutions has many subtleties, in particular concerning the
regularity of the metric \cite{bond,bon1,hel1}.
The purpose of this paper is to clarify, unify and complete
existing results on regularity. This topic has important implications,
for example in the exact modelling of gravitational collapse in an expanding universe
\cite{Matravers:2000cu}, or the exact modelling 
of cosmological voids in an expanding universe \cite{Sakai:2008fi,Grenon:2011fs}. 

The new results include the matching between
an exact solution in the generalised Kantowski-Sachs family and the LTB
solutions, and junction conditions for arbitrary non-null surfaces.
The possible types of centre are extended. (Note that spherically symmetric (SS) dust solutions
need not possess a centre \cite{hel2}.) {\it All} possible
composite SS dust models are found and
classified into four topological groups. One such topology, in which
spatial sections have the topology of a 3-torus, appears not to have
been discussed previously.

We use comoving spatial coordinates, since these are best adapted to
matching problems. For an analysis of astronomical observations,
coordinates based on the past light cones of the observer are a
better choice \cite{mhms}.

The paper is laid out as follows.
In section II, all of the solutions of Einstein's field equations (with
a smooth SS dust source) are expressed in the forms
(\ref{5sol}) and (\ref{6sol}). The geometrical requirements
at any junction in spacetimes composed from these solutions are
analysed in section III. In section IV,
the reasonable physical regularity conditions are made explicit.
This leads to regularity conditions within
the domain of any one solution (section V), at any centre of symmetry
(section VI), and at
any junction between solution domains (section VII). Also in section VII,
all the possible composite models are listed.
In section VIII, the results are summarised and used to give a simple
proof of the recollapse conjecture for these models.

\section*{II. Spherical Dust Solutions}

In standard comoving coordinates $x^a=\left\{t,r,\theta,\phi\right\}$,
the dust 4-velocity is $u^a=\delta^a_t$ and
the metric is \cite{bond}
\be
ds^{2}=-dt^{2}+X(r,t)^{2}dr^{2}+R(r,t)^{2}d\Omega^{2},
\label{metric}
\ee
where $d\Omega^{2}\equiv d\theta^{2}+\sin^{2}\theta\;d\phi^{2}$ and
we choose $R\geq 0$.
The energy-momentum tensor is
$T_{a}^{b}=\rho u_{a}u^{b}=-\rho \delta^{t}_{a}\delta^{b}_{t}$,
where $\rho$ is the proper matter energy density.
The Einstein tensor is \cite{bond}
\bea
G_{t}^{t}&=&-2\frac{\dot{X}\dot{R}}{XR}-\frac{1+\dot{R}^2}{R^{2}}
+\frac{1}{X^{2}}\left(2\frac{R^{\prime\prime}}{R}
+\frac{R^{\prime 2}}{R^{2}}-2\frac{X^{\prime}R^{\prime}}{XR}
\right), \label{b1}\\
G_{r}^{r}&=&-2\frac{\ddot{R}}{R}-\frac{1+\dot{R}^2}{R^{2}}
+\frac{R^{\prime 2}}{X^{2}R^{2}}, \label{b2}\\
G_{\theta}^{\theta}&=&G_{\phi}^{\phi}=-\frac{\ddot{X}}{X}
-\frac{\ddot{R}}{R}-\frac{\dot{X}\dot{R}}{XR}+\frac{1}{X^{2}}
\left(\frac{R^{\prime\prime}}{R}-\frac{X^{\prime}R^{\prime}}
{XR}\right), \label{b3}\\
G_{r}^{t}&=&X^{2}G_{t}^{r}=2\left(\frac{\dot{R}^{\prime}}{R}
-\frac{\dot{X}R^{\prime}}{XR}\right),\label{b4}
\eea
where an overdot denotes
$u^a\partial_a=\partial/\partial t$ and a prime denotes
$\partial/\partial r$. The units satisfy $G=c=1$.

In order for solutions of Einstein's equations
$G_{a}^{b}=8\pi T_{a}^{b}$ to exist, the Einstein tensor $G_{a}^{b}$ must
be defined through (\ref{b1})-(\ref{b4}).
In such regions of spacetime,
it is therefore required that
\be
X\neq 0\neq R; ~ X ~ {\rm and} ~ R ~ {\rm are} ~ C^2 ~ {\rm in}
~ t; ~ X ~ {\rm and} ~ \dot{R} ~ {\rm are} ~ C^1 ~ {\rm in} ~ r;
~ R ~ {\rm is} ~ C^2 ~
{\rm in} ~ r.\label{diffy}\ee
However, these regions may be joined together to form a composite
spacetime, in which $G_{a}^{b}$ is not defined by (\ref{b1}-\ref{b4})
on the boundaries.

Bondi \cite{bond} integrated the system as follows.
The $G_{t}^{r}=0$ field equation integrates to (assuming
$R^\prime\neq 0$)
\be X=\frac{R^{\prime}}{\sqrt{1+E(r)}},\label{int0}\ee where $E$ is an
arbitrary function. The remaining independent equations reduce to
\be \dot{R}^{2}=\frac{2M(r)}{R}+E, \label{oldsol}\ee with $M$ another
arbitrary function. The corresponding proper density is
given by\be \rho=\frac{M^{\prime}}{4\pi R^{\prime}R^2}.
\label{den1}\ee
There are five solutions of (\ref{oldsol}):
\be
\begin{array}{lllll}
{\rm \bf (s1)}&{\rm for}&\left\{E=M=0\right\}:&R=-T&[T\leq 0],\\
{\rm \bf (s2)}&{\rm for}&\left\{E>0,\;M=0\right\}:&
R=\tau\sqrt{E},~~\tau \equiv \epsilon t-T~~  \epsilon=\pm 1 & [\tau\geq 0],\\
{\rm \bf (s3)}&{\rm for}&\left\{E=0,\;M>0\right\}:&
R=\left(9M/2\right)^{1/3}\tau^{2/3} & [\tau\geq 0],\\
{\rm \bf (s4)}&{\rm for}&\left\{E>0,\;M>0\right\}:&
R=M\left(\cosh\eta-1\right)E^{-1},&
\sinh\eta-\eta=\tau E^{3/2}M^{-1}\\
&&&&[0<\eta<+\infty,~ \tau>0],\\
{\rm \bf (s5)}&{\rm for}&\left\{E<0,\;M>0\right\}:&
R=M\left(\cos\eta-1\right)E^{-1},&
\eta-\sin\eta=\tau|E|^{3/2}M^{-1}\\
&&&&[0<\eta<2\pi,~\tau>0],\end{array}\label{5sol}\ee
where $T(r)$ is a third arbitrary function and we denote the five solutions
by (s1),\dots, (s5).
No physical solutions exist
for $\left(E<0,\;M\leq 0\right)$ or for $\left(E=0,\;M<0
\right)$. Note
that (s1) and (s2) are (locally) Minkowski spacetime.
Motivated by equation (\ref{oldsol}),
Bondi \cite{bond} describes $M$ as a relativistic
generalisation of Newtonian mass, and $\frac{1}{2}E$ as a
total energy. The surfaces $\tau=0$ are spacelike singularities.

The above integration necessarily requires $R^{\prime}\neq 0$. If
instead $R^{\prime}=0$, then a different solution
results, as follows. The $G_{r}^{r}=0$ equation integrates
to
\be \dot{R}=\pm\sqrt{\frac{2\tilde{M}}{R}-1}, \label{int1}\ee
where $\tilde{M}>0$ is an arbitrary constant and
$\dot{R}\neq 0$ has been assumed, since the converse
immediately leads to an inconsistency.
The second integration
(assuming $0<R\leq 2\tilde{M}$, otherwise $\dot{R}^2<0$) reveals
that
\be R=\tilde{M}\left(1-\cos\eta\right),\;\;\;
\eta-\sin\eta=\tilde{M}^{-1}\left(
\epsilon t-\tilde{T}\right),\label{int3}\ee
where $0<\eta<2\pi$ and $\tilde{T}$ is an arbitrary constant.
By (\ref{int3}),
equation (\ref{b3}) implies the linear equation
\be \frac{\partial^2 X}{\partial \eta^2}+\frac{X}
{\cos\eta-1}=0,\label{int2}\ee
which transforms to a first-order Ricatti equation
under $X\rightarrow X^{-1}\partial_{\eta}X$.
Hence the general solution of (\ref{int2}) may be found
provided one solution is
known \cite{kam}. One particular solution is
$X=\sin\eta\left(1-\cos\eta\right)^{-1}$, and the
general solution is:
\be{\rm \bf (s6)}~~~~~~~~~~~~~X=A(r)\frac{\sin\eta}
{1-\cos\eta}+B(r)\left[1
-\frac{\eta\sin\eta}{2\left(1-\cos\eta\right)}
\right],\label{6sol}\ee
where $A$ and $B$ are arbitrary functions, and we denote this solution by
(s6).
Finally [by (\ref{b1})] the density reduces to
\be \rho=\frac{B}{8\pi\tilde{M}^2\left(1-\cos\eta\right)\left[
A\sin\eta+B\left(1-\cos\eta-\frac{1}{2}\eta
\sin\eta\right)\right]}.\label{den3}
\ee
This solution is an SS variant of an
inhomogeneous generalisation of the
$k=+1$ Kantowski-Sachs metric, as was discovered previously \cite{ell}.
The form of the solution presented in
\cite{kra} (in which $\rho\neq 0$ was assumed) is
\be ds^{2}=-e^{2\nu}d\tilde{t}^2+e^{2\lambda}d\tilde{r}^2
+\tilde{t}^2 d\Omega^2,~~
e^{2\nu}=\frac{\tilde{t}}{a-\tilde{t}},~~
e^{\lambda}=e^{-\nu}\left[
\int^{e^{\nu}}\frac{2x^2}{1+x^2}dx+C(\tilde{r})\right].\label{kra}
\ee
The coordinate transformation between (\ref{kra}) and (s6) is given by
$\tilde{t}=R$, $\tilde{r}^{\prime}=A/C$, $\dot{\tilde{r}}=0$ (with the
identifications $C=2A/B$ and $a=2\tilde{M}$).

In summary, the six possible SS dust solutions are
(s1)-(s6). Matching these dust solutions to form composite space times 
is a focus in the remaining sections. 

\section*{III. Junction Conditions}

The differentiability conditions (\ref{diffy}) on the metric need not be
satisfied at the interfaces
between domains of separate solutions in a composite
model. Instead (weaker) matching conditions (for the geometry) must
be satisfied \cite{isr}. At these junctions, it is assumed that the
coordinate $r$ remains regular, which is not the case at a
shell-crossing (by definition). Such singularities are excluded
in this paper.
The matching of two general SS spacetimes has been
considered in \cite{fay}, in which necessary
conditions for the matching were
presented, which are valid
for any equation of state. Here necessary and sufficient conditions
are found in the special case of dust.

Consider firstly the case of a comoving boundary
$r={\rm constant}$ (which was analysed previously \cite{bon1}).
The unit outward normal is $n_{a}=|X|\delta^{r}_{a}$,
the metric intrinsic to the surface is \be \hat{h}_{ab}=g_{ab}-n_{a}n_{b}
={\rm diag}\left(-1,0,R^2,R^2\sin^{2}\theta\right),\label{intmet}\ee
and the extrinsic curvature is
\be \hat{K}_{ab}=\hat{h}^{c}_{a}\hat{h}^{d}_{b}\nabla_{d}n_{c}
={\rm diag}\left(0,0,\frac{RR^{\prime}}{|X|},
\frac{RR^{\prime}}{|X|}\sin^{2}\theta\right).\label{extmet}\ee
The Darmois matching conditions \cite{bon1,isr,lak1}
state that $\hat{h}_{ab}$ must match across the surface
and any discontinuity in $\hat{K}_{ab}$
gives rise to a surface-density layer as represented by the
Lanczos equation \cite{lak1} \be 8\pi\left({}^3 T_{ab}\right)
=g^{cd}\left(\Delta K_{cd}\right)h_{ab}
-\Delta K_{ab},
\label{lanc}\ee
where $\Delta$ denotes the limit of a quantity on the `$r+$'
side of the interface, minus the limit on the `$r-$' side.
The surface energy-momentum tensor ${}^3T_{ab}$ can be expressed in perfect
fluid form
${}^3 T_{ab}=\left({}^3\rho+{}^{3}p\right)u_{a}u_{b}+{}^{3}p\hat{h}_{ab}$
with ${}^3\rho$ and ${}^{3}p$ the surface energy density and surface
(isotropic) pressure. Now
\[{}^3\rho={}^3 T_{ab}u^{a}u^{b}=-\frac{1}{4\pi R}\Delta\left(
\frac{R^{\prime}}{|X|}\right),~~~
{}^{3}p= {1\over2}\left( {}^3 T_{ab}\hat{h}^{ab} + {}^3\rho \right) =-\frac{{}^3\rho}{2}.\]
Following Bonnor \cite{bon1}, this equation of state
is regarded as unphysical, i.e. we require the
extrinsic curvature on comoving surfaces to be
continuous in $r$.
By (\ref{intmet}) and (\ref{extmet}), the junction conditions reduce to
\bea &&
R{\rm\;\;continuous
\;\;in\;\;}r,\label{mat1}\\
&& \frac{R^{\prime}}{|X|}
{\rm\;\;continuous\;\;in\;\;}r.
\label{mat2}\eea
For solution (s6), $R^\prime\equiv 0$, hence
(s6) may only be matched (across a comoving surface)
to one other SS dust solution, i.e. (s5)
[by (\ref{int0}), $\left\{R^\prime\rightarrow 0,~X\neq 0\right\}$
requires $E\rightarrow -1$].
From (\ref{5sol}), (\ref{int3}) and (\ref{mat1})
the matching also requires $M\rightarrow \tilde{M}$
and $T\rightarrow \tilde{T}$ in the (s5) region.
This motivates a characterisation of (s6):
{\it solution (s6) may be characterised within the LTB family by
the conditions $\left\{M^{\prime}=T^{\prime}=0,\;\;\;M>0,\;\;\;
E=-1\right\}$.}

As we show below in Section VII, in addition to (s5) to (s6),
the other possible matchings across a comoving boundary are: (s2) to (s4), (s3) to (s4) or (s5) 
and (s4) to (s5).

Consider now the junction conditions on the
spacelike surfaces $t={\rm constant}$. In this case,
the unit normal is $u^{a}$ and the
intrinsic and extrinsic curvatures are
\be h_{ab}=g_{ab}+u_{a}u_{b}
={\rm diag}\left(0,X^2,R^2,R^2\sin^{2}\theta\right),~~~
K_{ab}=h^{c}_{a}h^{d}_{b}\nabla_{d}u_{c}
={\rm diag}\left(0,\dot{X}{X},\dot{R}{R},\dot{R}{R}\right).
\label{extmet2}\ee
Hence continuity of $h_{ab}$ and $K_{ab}$ in $t$
merely implies that $R,X,\dot{R}$ and $\dot{X}$ are continuous in $t$.
(Spacelike surface layers, which imply an
instantaneous transition, are treated as unphysical a priori.)
In fact, by inspection of equations (\ref{int0}),
(\ref{oldsol}) and (\ref{int1}), the metric tensors
for (s1)-(s6)  are infinitely differentiable ($C^{\infty}$) in $t$.

An analysis of the Darmois matching problem for solutions (s1)-(s6)
across a general non-comoving (and non-null)
SS surface
provides insight into the nature of general
dust models. From (\ref{metric}), the unit normal $n^a$
and unit SS tangent $t^a$ to such a surface satisfy
\[ n_an^a=-t_at^a=\lambda,
~~~n_a=\left(-P,~|X|\sqrt{\lambda+P^2},~0,~0\right),
~~~t_a=\left(-\sqrt{\lambda+P^2},~P|X|,~0,~0\right),
\]
where the surface is timelike (spacelike) for $\lambda=1$
($\lambda=-1$) and $P$ is a function determined by the
equation of the surface:
\be r=s(t),~~~\frac{ds}{dt}=\frac{P}{|X|\sqrt{\lambda+P^2}}. \label{Peq}\ee
A lengthy calculation leads to the extrinsic curvature of
the surface:
\be
\hat{K}_{ab}=\left(
\begin{array}{cccc}
-\left(1+\lambda P^2\right)F_1 & F_1|X|\lambda P\sqrt{\lambda+P^2} & 0
& 0 \\
F_1|X|\lambda P\sqrt{\lambda+P^2} & -\lambda P^2X^2F_1 & 0 & 0 \\
0 & 0 & F_2 & 0 \\
0 & 0 & 0 & F_2\sin^2\theta\end{array}\right),\label{hatk}
\ee
and the intrinsic curvature of the surface is
\be \hat{h}_{ab}=\left(
\begin{array}{cccc}
-\left(1+\lambda P^2\right) & \lambda P|X|\sqrt{\lambda+P^2} & 0 & 0 \\
\lambda P|X|\sqrt{\lambda+P^2} & -\lambda P^2X^2 & 0 & 0 \\
0 & 0 & R^2 & 0 \\
0 & 0 & 0 & R^2\sin^2\theta\end{array}
\right),\label{hath}
\ee
where
\[F_1=\frac{P_{,a}t^a}{\sqrt{\lambda+P^2}}+P\frac{\dot{X}}{X},~~~
F_2=\frac{PR}{\sqrt{\lambda+P^2}}R_{,a}t^a+\frac{\lambda RR^\prime}
{|X|\sqrt{\lambda+P^2}}.\]
Since the coordinates continue through the surface,
all four coordinates are induced on the
surface, and one of $r,t$ is redundant there.
Now, the Darmois conditions with no surface layers take the form
\bea \Delta\hat{h}_{ab}=0~&\Rightarrow&~\Delta R=\Delta |X|=0,
\label{hatheq}\\
\Delta\hat{K}_{ab}=0~&\Rightarrow&~\Delta R^\prime=\Delta\big(
\dot{X}/X\big)=0,\label{hatkeq}\eea
where it has been assumed that $P\neq 0$ and $P\neq 1$, as these cases
are already considered above. By
equations (\ref{Peq}), (\ref{hath}) and (\ref{hatheq}),
$\Delta P=0$. We now show that:
\begin{quote}
{\it If there are no surface layers, then all boundaries
between domains of different solutions {\rm(s1)--(s6)}
in a composite dust model
must be comoving, i.e. $\left\{r={\rm constant}\right\}$.} 
\end{quote}
(See also \cite{new} and \cite{Grenon:2011fs}.)

The proof is as follows. In matching together two different
solutions (s1)-(s5), $|X|=|R^\prime|/\sqrt{1+E}$. Then by
(\ref{hatheq}) and (\ref{hatkeq}), $\Delta E=0$. From (\ref{Peq})
\[ \dot{R}=\frac{R_{,a}t^a}{\sqrt{\lambda+P^2}} 
-\frac{PR^\prime}{|X|\sqrt{\lambda+P^2}}, \] 
so that (\ref{hatheq}) and (\ref{hatkeq}) imply $\Delta\dot{R}=0$. Then by
(\ref{oldsol}), $\Delta M=0$. Finally $\Delta R=0$ forces
$\Delta T=0$. 
Note that $\Delta (R_{,a}t^a)=(\Delta R)_{,a}t^a=0$, since $t^a$ is tangent to the boundary. 
Hence all three LTB functions $E$, $M$ and $T$ `carry
through' the surface. Since $E$, $M$ and $T$ depend only on
$r$, the surface must be of the form $r={\rm const}$.
By (\ref{hatkeq}), matching between one of
(s1)-(s5) and solution (s6) requires $R^\prime\rightarrow 0$
on the (s1)-(s5) side.
This is only possible in (s5), with $E\rightarrow-1$.
However [by (\ref{int0})], $E(r)=-1$ only at isolated values of $r$,
i.e only on a comoving surface. The result follows.

It is of some interest however to establish the nature of
non-comoving singular surfaces (surface layers)
in these solutions, in which case
$\Delta \hat{h}_{ab}=0\neq\Delta \hat{K}_{ab}$.
Restricting to $\lambda=1$
(timelike surface layers) one may construct ${}^3T_{ab}$ once
again, by (\ref{lanc}):
\be 8\pi\left({}^3T_{ab}\right)=\left(\begin{array}{cccc}
-{2\sqrt{1+P^2}\Delta R^\prime}/{(R|X|)}&{2P\Delta R^\prime}/{R}
&0&0\\
{2P\Delta R^\prime}/{R}&-{2|X|P^2\Delta R^\prime}/{(R\sqrt{1+P^2})}
&0&0\\
0&0&-F_3&0\\
0&0&0&-F_3\sin^2\theta
\end{array}\right),\label{3energy}\ee
where \[F_3=\frac{R^2P^2\Delta\left(X^\prime/X\right)-R\Delta R^\prime}
{|X|\sqrt{1+P^2}}.\]
A comparison between (\ref{3energy}) and the perfect fluid
energy tensor
${}^3 T_{ab}=\left({}^3\rho+{}^3p\right)t_at_b+{}^3p\hat{h}_{ab}$
reveals that {\it the surface layer is always of perfect fluid form}.
The surface energy density and pressure are
\[ {}^3\rho=-\frac{\Delta R^\prime}{4\pi R|X|\sqrt{1+P^2}},
~~~{}^3p={}^3\rho\left(-\frac{1}{2}+\frac{P^2R\Delta\left(X^\prime/X
\right)}{2\Delta R^\prime}\right).\]
The nature of the matching problem
changes greatly in moving away from the comoving case.
In the non-comoving case, conditions at the wall
must be satisfied through some {\it range} of
$r$. These conditions are therefore `dragged'
into the adjoining
spacetimes, since the arbitrariness in these solutions resides purely
in functions of $r$.
This approach to surface layers and its application to models of
voids is further discussed in \cite{new}.\\

From now on,
singular surfaces are ruled out. All boundaries are necessarily
comoving, and the metrics
are fully determined then by choices of the arbitrary functions
$E(r)$, $M(r)$ and $T(r)$ [and
$A(r)$ and $B(r)$ in regions where $E(r)=-1$].
Throughout these SS dust models, junction condition (\ref{mat1})
reduces to
\be M(r)(\geq 0),\;E(r)(\geq -1)
{\rm\;\;and\;\;}T(r){\rm\;\;are\;\;continuous}.\label{mat3}\ee

\section*{IV. Regularity Requirements}

In this section, the physical requirements (and one coordinate
constraint)
to be imposed
on the metric are made explicit and justified. From section III, the
metrics of
(s1)-(s6) are $C^{\infty}$ in $t$, hence attention is
focused on
radial differentiability.

For a well-behaved radial coordinate, $g_{rr}$ must be piecewise continuous
in $r$ (that is, continuous except at isolated values
of $r$, where both left and right limits must be finite).
It is also
required that $\lim_{\pm}g_{rr}\neq 0$ everywhere, where a $+$ ($-$)
denotes a right (left) limit. This extra condition
purely defines a `good' spatial coordinate, so that
$dr$ is everywhere proportional to the differential increase in radial
proper distance. This subtlety ensures that the continuity properties of
physical quantities may be expressed unambiguously through their
differentiabilities in $r$.

The dust is characterised (in the SS case) by the
density $\rho$, expansion rate $\Theta=K_a{}^a$
and shear $\sigma_{ab}=K_{ab}-\frac{1}{3}\Theta h_{ab}$.
For physically reasonable matter, $\rho$, $\Theta$ and
$\sigma_{ab}$ must each be piecewise continuous in $r$.
By spherical symmetry,
$\sigma_{ab}\rightarrow 0$ wherever $R\rightarrow 0$.
For the metric (\ref{metric})
\be
\sigma_{a}^{b}=\frac{1}{3}\left[\frac{\dot{X}}{X}-\frac{\dot{R}}{R}
\right]\times
{\rm diag}\left(0,2,-1,-1\right),
~~~\Theta=2\frac{\dot{R}}{R}+\frac{\dot{X}}{X},\label{theta}\ee
where $\dot{X}/X$ and $\dot{R}/R$ are `radial' and `azimuthal'
expansion rates.

To ensure that the spacetime itself is regular at each point
\cite{mtw}
it is required that\\
$R_{(i)(j)(k)(l)}
=R_{abcd}e_{(i)}^{a}e_{(j)}^{b}e_{(k)}^{c}e_{(l)}^{d}$ is piecewise
continuous in $r$, where
$e_{(i)}^{a}$ is an orthonormal tetrad basis. Here a natural
choice is made:
$e_{(0)}^{a}=u^{a}$, $e_{(1)}^{a}=|X|^{-1}\delta^{a}_{r}$,
$e_{(2)}^{a}=R^{-1}\delta^{a}_{\theta}$,
$e_{(3)}^{a}=R^{-1}{\rm cosec}\theta
\delta^{a}_{\phi}$. For solutions (s1)-(s5), the nontrivial
components are found to be
\bea
&&R_{(0)(1)(0)(1)}=4\pi\rho-\frac{2M}{R^3},~~~
R_{(0)(2)(0)(2)}=\frac{M}{R^3},\nonumber\\
&&R_{(1)(2)(1)(2)}=4\pi\rho-\frac{M}{R^3},~~~
R_{(2)(3)(2)(3)}=\frac{2M}{R^3}.\nonumber
\eea
and for (s6)
\bea
&&R_{(0)(1)(0)(1)}=\frac{\dot{X}\dot{R}}{XR}-\frac{M}{R^3},~~~
R_{(0)(2)(0)(2)}=\frac{M}{R^3},\nonumber\\
&&R_{(1)(2)(1)(2)}=\frac{\dot{X}\dot{R}}{XR},~~~
R_{(2)(3)(2)(3)}=\frac{2M}{R^3}.\nonumber
\eea
Hence, for regular spacetimes: at points where $R\rightarrow 0$,
$\lim (M/R^3)$ must be finite [using (\ref{mat3}), and since $\rho$,
$\dot{X}/X$ and $\dot{R}/R$ are
already required to be piecewise continuous in $r$].
Solution (s6) does not admit central
points. To summarise, throughout the models
it is required that:

\begin{itemize}
\item[{\bf R1.~}] The junction condition (\ref{mat2}) is satisified,

\item[{\bf R2.}~] $g_{rr}$ is piecewise continuous in $r$ and
$\lim_{\pm}g_{rr}\neq 0$,

\item[{\bf R3.}~] $\rho\geq 0$ is piecewise continuous in $r$,

\item[{\bf R4.}~] $\dot{R}/R$ is piecewise continuous in $r$,

\item[{\bf R5.}~] $\dot{X}/X$ is piecewise continuous in $r$, and
$\dot{X}/X\rightarrow\dot{R}/R$ wherever $R\rightarrow 0$,

\item[{\bf R6.}~] $M/R^3$ is finite wherever $R\rightarrow 0$,
\end{itemize}
(except, trivially, at the spacelike singularities $\tau\rightarrow 0$).

In sections V-VII the above conditions
[with (\ref{mat3}) satisfied a priori]
are enforced for a general SS dust metric,
to guarantee regularity.

\section*{V. Regular Solutions}

Here the conditions of section IV are verified in turn, for points
{\it in the domain} of a solution. This domain does {\it not}
include the origin (treated in section VI)
or interfaces between solutions (treated in section VII).
\[\]
{\noindent{\bf R1.}} Junction condition (\ref{mat2}) is automatically
satisfied for (s6).
For any of (s1)-(s5) it implies that [using (\ref{int0})]
\be R^{\prime}~{\rm\;\;may
\;\;change\;\;
sign\;\;only\;\;at\;\;values\;\;of\;\;}r{\rm\;\;satisfying
\;\;}E(r)=-1,\label{mat4}\ee
as was noted in \cite{bon1}.
Hence $R^{\prime}\geq 0$
throughout the domain of each solution (s1)-(s4),
or $R^{\prime}\leq 0$ throughout. In (s5),
$R^\prime$ may change sign where $E=-1$. Now
for (s4), $R^{\prime}$ may be written as
\be R^{\prime}=\frac{M^{\prime}}{E}\left[
\frac{\eta\sinh\eta}{\cosh\eta-1}-2\right]-T^{\prime}E^{1/2}\left[
\frac{\sinh\eta}{\cosh\eta-1}\right]+\frac{E^{\prime}M}{E^2}
\left[\frac{\sinh\eta\left(\sinh\eta-3\eta\right)}{2\left(\cosh\eta
-1\right)}+2\right],\label{rp4}
\ee
and for a positive density [by equation (\ref{den1})], $R^{\prime}$ and
$M^{\prime}$ must have the same sign. At large $\eta$ (large $t$),
the third term
in (\ref{rp4}) dominates, so that $R^{\prime}$ and $E^{\prime}$ must have
the same sign. At small $\eta$, the second term dominates, so
$T^{\prime}$ must have the opposite sign to $R^{\prime}$. These
conditions are also sufficient (to ensure $R^\prime$ has a constant
sign), since in (\ref{rp4}) each function of
$\eta$ in square brackets is strictly positive for all allowed $\eta$.
Hence [for (s4)] equation (\ref{mat4}) implies that
\be \pm R^{\prime}\geq 0\Rightarrow \left\{\pm M^{\prime}\geq 0,
\;\;\;\pm E^{\prime}\geq 0,\;\;\;\pm T^{\prime}\leq 0\right\}.
\label{1234conds}\ee
For any of (s1)-(s3), similar reasoning also leads to (\ref{1234conds}).
Now for regions of (s5) which satisfy
$E\neq -1$ (in which $R^{\prime}$ cannot change
sign), it is useful to write $R^\prime$ in the form
\bea R^{\prime}&=&\left(\frac{E^{\prime}M}{E^2}-\frac{2M^{\prime}}
{3E}+\frac{T^{\prime}|E|^{1/2}}{3\pi}\right)
\left[\frac{\sin\eta
\left(\sin\eta-3\eta\right)}{2\left(1-\cos\eta\right)}+2\right]\nonumber\\
&&-T^{\prime}|E|^{1/2}\left[\frac{\sin\eta}{1-\cos\eta}
+\frac{1}{3\pi}\left\{2+\frac{\sin\eta\left(\sin\eta-3\eta\right)}
{2\left(1-\cos\eta\right)}\right\}\right]
+\frac{M^{\prime}}{3|E|}\left[1-\cos\eta\right],\label{rp5}\eea
where each function in square brackets is always
positive. At small
$\eta$ the second term dominates. As $\eta\rightarrow 2\pi$ the first term
dominates, and $M^{\prime}$ must have the same sign as $R^{\prime}$
for $\rho\geq 0$. Hence in these regions of (s5)
\be \pm R^{\prime}\geq 0\Rightarrow \left\{\pm M^{\prime}\geq 0,
\;\;\;\pm\left(E^{\prime}M-\frac{2}{3}M^{\prime}E
+\frac{T^{\prime}|E|^{5/2}}{3\pi}\right)\geq 0,\;\;\;\pm T^{\prime}\leq 0
\right\}.\label{5conds}\ee
Equations (\ref{1234conds}) and (\ref{5conds}) are
the Hellaby and Lake no-shell-crossing conditions \cite{hel1},
derived here from the junction conditions.
Violation of (\ref{1234conds}) or (\ref{5conds}) would necessitate
either a pathological choice of radial coordinate or true caustic
formation.
\[\]
{\noindent{\bf R2.}} From (\ref{int0}), piecewise continuity of the radial
coordinate
within the domain of one of (s1)-(s4) [and in regions of (s5) where
$E(r)\neq-1$] requires piecewise continuity of $R^{\prime}$ in
$r$, for which it is necessary and sufficient that
$M^{\prime}(r)$,
$T^{\prime}(r)$ and $E^{\prime}(r)$ are piecewise continuous.
Now $\lim_{\pm}g_{rr}\neq 0$ implies
$\lim_{\pm} R^{\prime}\neq 0$ in these regions, which
[from (\ref{1234conds}) and (\ref{5conds})] forces one of
$\lim_+ M^{\prime}$, $\lim_+ T^{\prime}$ and
$\lim_+ E^{\prime}$ to be nonzero at each point (likewise for the
left limits). At points in (s5)
where $E(r)=-1$, $\lim_+g_{rr}$ is finite if and only if
\be \lim{{}_+}\frac{M^\prime}{\sqrt{1+E}},\;\;\;\lim{{}_+}\frac{T^\prime}
{\sqrt{1+E}}\;\;\;{\rm and}\;\;\;\lim{{}_+}\frac{E^\prime}{\sqrt{1+E}}
\;\;\;{\rm are\;finite,\;and\;at\;least\;one\;is\;nonzero},
\label{e=-1}\ee
as follows from (\ref{int0}), (\ref{rp5}) and (\ref{5conds}) (and
analogously for $\lim_-g_{rr}$).
One necessary consequence of (\ref{e=-1}) is that
$\lim_{\pm}R^{\prime}=\lim_{\pm}M^{\prime}=\lim_{\pm}T^{\prime}
=\lim_{\pm}E^{\prime}=0$ at these points.
Finally, for (s6) the radial coordinate is
well-behaved if $A(r)$ and $B(r)$ are piecewise continuous and if at
least one of
$\lim_{+}A$ and $\lim_{+}B$ is nonzero [by (\ref{6sol}), and likewise for
$\lim_-$].
\[\]
{\noindent{\bf R3.}} The density is piecewise
continuous in (s1)-(s5) as a consequence of the junction conditions,
which is seen as follows.
From (\ref{den1}), $\rho$ is
at least continuous (in $r$) except at isolated points,
since $R^{\prime}$ and $M^{\prime}$ have this property (see above)
and $R^\prime=0$ only at isolated points.
Trivially $\rho=0$ for
(s1) and (s2). For (s3)-(s5),
if $M^{\prime}=0$ in a finite region, then $\rho=0$ there.
Otherwise $\lim_{\pm}\rho$ are finite if and only if
$\lim_{\pm}(R^\prime/M^\prime)\neq 0$, but this is automatically
satisfied in (s3)-(s5)
[e.g. in (s4)
\[\lim{{}_\pm}\frac{R^\prime}{M^\prime}\geq\frac{1}{E}\left[
\frac{\eta\sinh\eta}{\cosh\eta-1}-2\right]>0,\]
by (\ref{rp4}),(\ref{1234conds})].
Hence $\lim_{\pm}\rho$ are finite, and $\rho$
is piecewise continuous in $r$. In fact, $\lim_{\pm}\rho=0$
(with $M^{\prime}\neq 0$) only if either $\lim_{\pm}(E^{\prime}
/M^{\prime})=+\infty$ or $\lim_{\pm}(T^{\prime}/M^{\prime})
=-\infty$. Note that none of these remarks require any modification at
points in (s5) with $E=-1$.
Now the density in (s6) [given by
(\ref{den3})] vanishes if $B=0$ and $A\neq 0$ [and (s6)
degenerates to part of the exterior Schwarzschild solution].
Otherwise $\rho$ is finite and positive at all times in (s6)
if and only if
\be \lim{{}_{\pm}}\frac{B}{A}\geq\frac{1}{\pi},\label{newsolcond}\ee
which ensures no zeroes in the denominator of (\ref{den3}).
Piecewise continuity of $\rho$ in (s6) follows from the piecewise
continuity of $A(r)$ and of $B(r)$.
\[\]
{\noindent{\bf R4.}} Full continuity of $\dot{R}/R$ in $r$ is guaranteed
throughout the SS dust spacetimes by (\ref{oldsol}) and
(\ref{mat3}).
\[\]
{\noindent{\bf R5.}} For (s1), $\dot{X}/X=0$. For (s2)-(s5) [by
(\ref{int0})-(\ref{den1})]
\[ \frac{\dot{X}}{X}=\frac{R}{\dot{R}}\left(4\pi\rho-\frac{M}{R^3}
+\frac{E^{\prime}}{2R^{\prime}R}\right).
\]
Hence piecewise continuity of the shear in (s1)-(s5) requires the
further condition that
$\lim_{\pm}(E^\prime/R^\prime)$ is finite wherever $R^\prime\rightarrow 0$.
This is trivially satisfied in (s3). In (s2), (s4) and (s5) it is also
automatically satisfied since $\lim_\pm(R^\prime/|E^\prime|)>0$ [e.g.
in (s4) \[ \lim{{}_\pm}\frac{R^\prime}{E^\prime}\geq
\frac{M}{E^2}\left[\frac{\sinh\eta\left(\sinh\eta-3\eta\right)}
{2\left(\cosh\eta-1\right)}+2\right]>0\]
by (\ref{rp4}),(\ref{1234conds})].
Finally, solution (s6) has
\[ \frac{\dot X}{X}=\frac{\epsilon}{M\left(1-\cos\eta\right)}\left[
\frac{\frac{1}{2}B\left(
\eta-\sin\eta\right)-A}{B\left(1-\cos\eta-
\frac{1}{2}\eta\sin\eta\right)+A\sin\eta}\right],
\]
which is automatically piecewise continuous in $r$ by
(\ref{newsolcond}) and by the piecewise continuity of $A(r)$ and $B(r)$.

\section*{VI. Regular Centres}

In this section the possible types of origin (for which $R\rightarrow 0$)
are determined by imposing
the conditions of section IV. Only comoving origins are possible, and
they may join only to solutions (s1)-(s5).
All the results are given in Table \ref{centraltab}, in which
(i) derives from the condition $R\rightarrow 0$,
(ii) derives from $\lim_\pm g_{rr}\neq 0$, and (iii) forces the shear to
vanish. In each case $M/R^3\rightarrow \frac{4}{3}\pi\rho$, which is
the Newtonian limit. Examples are provided in the Table; in each the origin lies
at $r=0$. The allowed ranges of $\tau, \eta$ follow from \eqref{5sol}. The central behaviours of (s4) and (s5)
generalize previous results. 
\begin{table}
\begin{center}
\begin{tabular}{|l|cl|l|l|}\hline
Soln. &      & Behaviour of $E$, $M$ and $T$   & Kinematics & Example\\
\hline\hline
         & & & & \\
         & (i)  & $T\rightarrow 0$                & $\Theta\equiv 0$ & \\
(s1)     &      &                                 & & $T=-r$\\
         & (ii)  & $\lim T^\prime$ finite, nonzero & $\rho\equiv 0$ &\\
         &      &                                 && \\
\hline
         & & & &\\
         & (i)  & $E\rightarrow 0$   & $\Theta\rightarrow 3\epsilon\tau^{-1}$ & $E=r^2$\\
(s2)     & (ii)  & $\lim (E^{-1/2}E^\prime)$ finite, nonzero & &\\
         & (iii)  & $ET^\prime/E^\prime\rightarrow 0$  & $\rho\equiv 0$ & $T=0$\\
         &      &                                  & &\\
\hline
         & & & &\\
         & (i)  & $M\rightarrow 0$  & $\Theta\rightarrow 2\epsilon\tau^{-1}$ & $M=r^3$\\
(s3)     & (ii)  & $\lim (M^{-2/3}M^\prime)$ finite, nonzero & & \\
         & (iii)  & $MT^\prime/M^\prime\rightarrow 0$ & $4\pi\rho\rightarrow\frac{2}{3}\tau^{-2}$ & $T=0$\\
         &      &     & & \\
\hline
         & & & & \\
         & (i)  & $E^{3/2}/M\rightarrow 0$, $M\rightarrow 0$ & $\Theta\rightarrow 2\epsilon\tau^{-1}$ & $E=r^3$\\
         & (ii)  & $\lim (M^{-2/3}M^\prime)$ finite, nonzero  & & $M=r^3$\\
         & (iii)  & $\lim (MT^\prime/M^\prime)=\lim(M^{1/3}E^\prime/M^\prime)=0$   & $4\pi\rho\rightarrow\frac{2}{3}\tau^{-2}$ & $T=0$\\
         &      &  & &\\
\cline{2-5}
         & & & &\\
         & (i)  & $E^{3/2}/M\rightarrow+\infty$, $E\rightarrow 0$ & $\Theta\rightarrow 3\epsilon\tau^{-1}$ & $E=r^2$\\
(s4)     & (ii)  & $\lim(E^{-1/2}E^\prime)$ finite, nonzero & & $M=r^4$\\
         & (iii)  & $\lim(ET^\prime/E^\prime)$ & $4\pi\rho\rightarrow 0$ & $T=0$\\
         &      & $=\lim\left[E^{-1/2}M^\prime E^{\prime -1}\log(E^{3/2}M^{-1})\right]=0$  & &\\
\cline{2-5}
         & & & &\\
         & (i)  & $E^{3/2}/M\rightarrow\alpha>0$, $M\rightarrow 0$ & $\Theta\rightarrow 3\epsilon\alpha\sinh\eta/(\cosh\eta-1)^2$ & $E=r^2$\\
         & (ii)  & $\lim(E^{-1}M^\prime)$ finite, nonzero & & $M=r^3$\\
         & (iii)  & $E^{-1}ME^\prime/M^\prime\rightarrow 2/3$ and $MT^\prime/M^\prime\rightarrow 0$  & $4\pi\rho\rightarrow 3\alpha^2/(\cosh\eta-1)^3$ & $T=0$ \\
         &      &  & &\\
\hline
         & & & &\\
         & (i)  & $|E|^{3/2}/M\rightarrow 0$, $M\rightarrow 0$ & $\Theta\rightarrow 2\epsilon\tau^{-1}$ & $E=-r^3$\\
         & (ii)  & $\lim(M^{-2/3}M^\prime)$ finite, nonzero & & $M=r^3$\\
         & (iii)  &  $\lim(MT^\prime/M^\prime)=\lim(M^{1/3}E^\prime/M^\prime)=0$& $4\pi\rho\rightarrow\frac{2}{3}\tau^{-2}$ & $T=0$\\
(s5)     &      &  & &\\
\cline{2-5}
         & & & &\\
         & (i)  & $|E|^{3/2}/M\rightarrow\alpha>0$, $M\rightarrow 0$ & $\Theta\rightarrow 3\epsilon\alpha\sin\eta/(1-\cos\eta)^2$ & $E=-r^2$\\
         & (ii)  & $\lim(E^{-1}M^\prime)$ finite, nonzero & & $M=r^3$\\
         & (iii)  & $E^{-1}ME^\prime/M^\prime\rightarrow 2/3$ and $MT^\prime/M^\prime\rightarrow 0$ & $4\pi\rho\rightarrow 3\alpha^2/(1-\cos\eta)^3$ & $T=0$\\
         &      &    & &\\
\hline
\end{tabular}
\caption{Central behaviour for regular centres. (i) follows from $R\rightarrow 0$,
(ii) from $\lim_\pm g_{rr}\neq 0$, and (iii) from $\sigma_{ab}\to 0$. }\label{centraltab}
\end{center}
\end{table}

One central limit suffices to illustrate the arguments used to obtain
Table \ref{centraltab}. At an origin of (s4),
suppose $\eta\rightarrow\infty$.
Then by (\ref{5sol}), on approaching the origin
\be e^\eta\approx\frac{2E^{3/2}\tau}{M}\rightarrow\infty,\label{eta2}\ee
and $R\rightarrow 0$ forces $E\rightarrow 0$, $M\rightarrow 0$.
Now
\[
\frac{\dot{R}}{R}\rightarrow\frac{\epsilon}{\tau},~~~
\frac{\dot{X}}{X}\rightarrow
\frac{\epsilon\left(\frac{1}{2}E^\prime E^{-1/2}+T^\prime E^{-1}
M\tau^{-2}+M^\prime E^{-1}\tau^{-1}\right)}
{\left[M^\prime E^{-1}\log\left(E^{3/2}M^{-1}\right)-T^\prime E^{1/2}
+\frac{1}{2}E^{\prime}E^{-1/2}\tau\right]},\]
so that vanishing shear requires either
\be \lim \left(\frac{T^\prime M}{E^{1/2}E^\prime}\right)=\lim\left(
\frac{M^\prime}{E^{1/2}E^\prime}\right)
=\lim\left[-\frac{T^\prime E}{E^\prime}+\frac{M^\prime}{E^{1/2}E^\prime}
\log{\left(\frac{E^{3/2}}{M}\right)}\right]=0\label{poss1}\ee
or
\be \lim\left(\frac{E^{1/2}E^\prime}{M^\prime}\right)
=\lim\left(\frac{MT^\prime}{M^\prime}\right)=0,
~~~\lim\left[\log\left(\frac{E^{3/2}}{M}\right)-
\frac{T^\prime E^{3/2}}{M^\prime}\right]=1.
\label{poss2}\ee
However the latter case (\ref{poss2}) is ruled out since the
logarithmic term must diverge, by (\ref{eta2}).
Hence (\ref{poss1}) is the only possibility, and this
reduces to \be\lim\left(\frac{T^\prime E}{E^\prime}\right)
=\lim\left[\frac{M^\prime}{E^{1/2}E^\prime}\log\left(
\frac{E^{3/2}}{M}\right)\right]=0,\label{2.57}\ee
by consideration of
(\ref{1234conds}) and (\ref{eta2}).
From (\ref{rp4})
\[ \sqrt{g_{rr}}\rightarrow R^{\prime}\rightarrow \frac{M^\prime}{E}
\log\left(\frac{E^{3/2}}{M}\right)-T^{\prime}E^{1/2}
+\frac{E^\prime\tau}{2E^{1/2}},\]
so that [with (\ref{2.57})], $\lim g_{rr}\neq 0$ requires
$\lim \left(E^\prime E^{-1/2}\right)$ finite and nonzero.
Finally
\be 4\pi\rho\rightarrow \left[\tau^2\log\left(\frac{E^{3/2}}{M}\right)
-\frac{E^{3/2}T^\prime\tau^2}{M^\prime}+\frac{E^{1/2}E^\prime\tau^3}
{2M^\prime}\right]^{-1},\label{den41}\ee
which vanishes (again by the divergence of the logarithmic term).

\section*{VII. Regular Interfaces}

In this section the conditions of section IV are considered on
the comoving interfaces between domains of solutions (s1)-(s6) in
a composite model. The solution domains are assumed to be regular
(in the sense of section V) and this generally ensures
regular interfaces.

By (\ref{mat4}), the sign of $R^{\prime}$ cannot change across
these interfaces, since $E\neq-1$ on them [except on interfaces
between (s5) and (s6), but $R^{\prime}\equiv 0$ in (s6)].
Solution (s1) may not be matched to any other solution, since
$\dot{R}\equiv 0$ in (s1) (whereas the other solutions are cosmological).
Solution (s2) does not match to (s3), because $M\rightarrow 0$
forces $R\rightarrow 0$ in (s3). Also, (s2) does not match to (s5), since
($E\rightarrow 0$, $M\rightarrow 0$) forces $R\rightarrow 0$ in
(s5). From
section III, (s6) only matches to (s5).
There remain just
five physical types of junction, given below. At each of the five, equation
(\ref{mat3}) ensures continuity of $R$.
\[\]
{\bf a. Matching (s2) to (s4)}\\
The (s2) side of this interface is unconstrained by the matching.
On approaching the interface from (s4), $M\rightarrow 0$, $E\neq 0$
and $M^\prime>0$ throughout some finite region [by the piecewise
continuity of $M^\prime$, and since $M>0$ in (s4)]. Hence
by (\ref{mat4}) and (\ref{1234conds}),
$R$ must increase in the direction (s2)$\to$(s4). Now
$\eta$ obeys (\ref{eta2}), so that in (s4)
\[\sqrt{g_{rr}}\rightarrow\frac{1}{\sqrt{1+E}}
\left[\frac{M^\prime}{E}
\log\left(\frac{E^{3/2}}{M}\right)-T^\prime E^{1/2}+\frac{E^\prime\tau}
{2E^{1/2}}\right].\]
Therefore $r$ is a good coordinate if
$\lim_{\rm(s4)} T^{\prime}$, $\lim_{\rm(s4)} E^{\prime}$ and
$\lim_{\rm(s4)}M^{\prime}\log M$ are finite, and at least one
of them is nonzero.

On the (s4) side, the density
reduces to (\ref{den41}), and vanishes by the divergence of the
logarithmic term. [Note that $\rho\equiv 0$ in (s2).]

On the (s4) side, since ($M\rightarrow 0$, $M^\prime\log M$ finite)
forces $M^\prime\rightarrow 0$, we have
\[ \frac{\dot{X}}{X}\rightarrow\left\{
\begin{array}{ll}
0 & {\rm if}~E^\prime\rightarrow 0,\\
\epsilon\left[\tau-{2ET^\prime}/{E^\prime}
-{2M^\prime}\log M/ ({E^{1/2}E^\prime})\right]^{-1} & {\rm otherwise},
\end{array}\right.\]
whereas on the (s2) side
\[\frac{\dot{X}}{X}\rightarrow\frac{\epsilon}{\tau-2ET^\prime/E^\prime},\]
and on both sides $\dot{R}/R\rightarrow\epsilon/\tau$. Hence the shear
is necessarily finite on both sides of the interface, as required.
\[\]
{\bf b. Matching (s3) to (s4)}\\
The (s3) side is unconstrained by the matching. On approaching
the interface from
(s4), $E\rightarrow 0$ while $M\neq 0$, so that
$\eta\approx (6\tau/M)^{1/3}E^{1/2}\rightarrow 0$,
and $E^\prime>0$ throughout some finite region [since $E>0$ in (s4)]. Hence
by (\ref{mat4}) and (\ref{1234conds}), $R$ must increase in the
direction (s3)$\to$(s4). On the (s4) side
\be \sqrt{g_{rr}}\rightarrow M^\prime\left(\frac{\tau^2}{6M^2}\right)^{1/3}
-T^\prime\left(\frac{4M}{3\tau}\right)^{1/3}+\frac{E^\prime}{40}
\left[\frac{\left(6\tau\right)^4}{M}\right]^{1/3},\label{pft1}\ee
so that $r$ is a good coordinate provided
\be \lim{}_{\rm(s4)}M^{\prime},~
\lim{}_{\rm(s4)}T^{\prime}~{\rm and}
~\lim{}_{\rm(s4)}E^{\prime}~{\rm are~finite,~
and~at~least~one~is~nonzero}.\label{pft2}\ee
On both sides of the interface, the density reduces to
\be 4\pi\rho\rightarrow\left\{\begin{array}{ll}
0&{\rm if}~M^\prime\rightarrow 0,\\
\left[3\tau^2/2-{3MT^\prime\tau}/{M^\prime}
+{\left(6\tau\right)^{8/3}M^{1/3}E^\prime}/({160M^\prime})
\right]^{-1} & {\rm otherwise},\end{array}\right.\label{pft3}\ee
and on both sides,
\be \frac{\dot{X}}{X}\rightarrow
\frac{\epsilon\left[T^\prime+{M^\prime\tau}/{M}
+\left\{ {243\tau^5}/{(250M^2)}\right\}^{1/3} E^\prime
\right]} {\left[{3M^\prime\tau^2}/{(2M)} -3T^\prime\tau
+{E^\prime}\left({9\tau^2}/{2}\right)^{4/3}/ {(10M^{2/3})}\right]}
,~~~~~~~~~~~~\frac{\dot{R}}{R}\rightarrow\frac{2\epsilon}{3\tau}.
\label{pft4}\ee
Therefore both $\rho$ and $\dot{X}/X$ are well-behaved.
\[\]
{\bf c. Matching (s3) to (s5)}\\
This interface is similar to {\bf b}.
On approaching the interface in (s5),
$\eta\approx (6\tau/M)^{1/3}|E|^{1/2}\rightarrow 0$.
Then all the results in {\bf b} for the kinematics and radial coordinate
(\ref{pft1})-(\ref{pft4}) follow [with `(s4)' replaced by `(s5)'].

However, $R$ must increase in the {\it opposite
sense} to that in {\bf b} - here $R$ must increase in the direction
(s5)$\to$(s3), as we now show.
Taking the limit of equation (\ref{5conds}) on the
(s5) side gives
${\rm sign}(R^\prime)E^\prime>0$. Since $E<0$ in (s5) (and $E^\prime$ is
piecewise continuous), the result follows.
Note that the coordinate condition (\ref{pft2}) is crucial to
this proof.
Note also that since $\dot{R}/R\rightarrow \frac{2}{3}\epsilon
\tau^{-1}>0$ at any interface between (s3) and (s5), the $r$-continuity
of $\dot{R}/R$ forces the existence of a finite region in (s5)
adjoining the interface for which the azimuthal expansion rate is
positive $\dot{R}/R>0$ [even though all points in (s5)
eventually satisfy $\dot{R}<0$].
\[\]
{\bf d. Matching (s4) to (s5)}\\
Both sides are constrained by $E\rightarrow 0$
with $M>0$, and the resulting junction is given precisely by combining
the results of {\bf b} and {\bf c}. In this case, $R$ must increase
in the direction
(s5)$\to$(s4), by a similar argument to that given in {\bf b.}
\[\]
{\bf e. Matching (s5) to (s6)}\\
At this interface the (s6) side is unconstrained by the matching.
Equation (\ref{mat4}) places no restriction, and $R$ may increase
in either direction on approaching the interface from (s5).
Now $\lim_{\rm(s5)}E=-1$. Therefore, as discussed
in section V, $r$ is a good coordinate
provided (\ref{e=-1}) is satisfied.

The (s5)-limits of $\rho$ and $\dot{X}/X$ are just those of an
`ordinary' point, i.e. one in the domain of (s5).
In this sense, the matching conditions at this type of
interface are considerably less restrictive than those at the other four.
\[\]
Combining these results with the rest of the paper, {\it all} the
regular SS dust models may be classified into four topologies:
\[\]
{\bf i. Open models with one origin}\\
By noting the sense in which $R$ must increase at the interfaces
{\bf a}-{\bf e} above, the only possible composite models are:
\[\begin{array}{ll}
{\cal O}{\rm(s1)}^+,    & {\cal O}{\rm(s2)}^+{\rm(s4)}^+,\\
{\cal O}{\rm(s2)}^+,    & {\cal O}{\rm(s3)}^+{\rm(s4)}^+,\\
{\cal O}{\rm(s3)}^+,    & {\cal O}{\rm(s5)}^+{\cal S}{\rm (s5)}^+{\rm(s3)}^+,\\
{\cal O}{\rm(s4)}^+,    & {\cal O}{\rm(s5)}^+{\cal S}{\rm (s5)}^+{\rm(s4)}^+,\\
{\cal O}{\rm(s5)}^+{\cal S},~~~~~~~~~~ &
{\cal O}{\rm(s5)}^+{\cal S}{\rm (s5)}^+{\rm(s3)}^+{\rm(s4)}^+,\end{array}\]
where ${\cal O}$ denotes an origin, and a superscript $+$ ($-$) implies
that $R$ increases (decreases) from left to right. Here ${\cal S}$
is {\it any} combination of ${\rm(s5)}^-$,
${\rm(s5)}^+$ and ${\rm(s6)}$.
Note that open models can be constructed from collapsing solutions
[e.g. ${\cal O}{\rm(s5)}^+{\rm(s6)}$]. Papapetrou \cite{pap} discussed a
particular example of ${\cal O}{\rm(s5)}^+{\rm(s3)}$.

In the above construction, we have noted from (\ref{int0}) that on
$t=$const, $d\chi=|dR|/\sqrt{1+E}$, where $\chi$ is radial proper
distance. Hence by (\ref{mat4}), if $E>\alpha>-1$ for all
$\chi>\beta$ ($\alpha,\beta$ constants) then:
\be\chi\rightarrow\infty~~{\rm forces}~~R\rightarrow\infty~~~~~
{\rm if}~~\frac{dR}{d\chi}>0,\label{asymp1}\ee
\be{\rm there~is~a~finite~value~of~}\chi>\beta~{\rm~for~which~}
R=0,~~~~~{\rm if}~\frac{dR}{d\chi}<0.\label{asymp2}\ee
However, if $E\rightarrow-1$ as $\chi\rightarrow\infty$, then neither
of (\ref{asymp1}),({\ref{asymp2}) are necessary. An example of
${\cal O}{\rm(s5)}^+{\rm(s5)}^-$ of type {\bf i} is
\bea
&&E=\left\{\begin{array}{lll}
-{\sin^2r}\left[1-{\rm e}^{-2r_0}\right]/{\sin^2r_0} &{\rm for}&0<r<r_0,\\
-1+{\rm e}^{-2r}&{\rm for}&r>r_0,
\end{array}\right.~~~
M=\left\{\begin{array}{lll}
{\sin^3r}\left[M_\infty+{\rm e}^{-r_0}\right]/{\sin^3r_0}&{\rm for}&
0<r<r_0,\\
M_\infty+{\rm e}^{-r}&{\rm for}&r>r_0,
\end{array}\right.\nonumber\\
&&T=0,~~~M_\infty>0,~~~\pi<r_0<2\pi,
\label{example1}\eea
and
\bea
&&E=\left\{\begin{array}{lll}
-{r^2}\left[1-{\rm e}^{-2r_0}\right]/{r_0^2}&{\rm for}&0<r<r_0,\\
-1+{\rm e}^{-2r}&{\rm for}&r>r_0,
\end{array}\right.~~~
M=\left\{\begin{array}{lll}
{r^3}\left[M_\infty-{\rm e}^{-r_0}\right]/{r_0^3} &{\rm for}&
0<r<r_0,\\
M_\infty-{\rm e}^{-r}&{\rm for}&r>r_0,
\end{array}\right.\nonumber\\
&&T=0,~~~0<M_\infty<2/3,~~~r_0>0,
\label{example2}\eea
is an example of ${\cal O}{\rm(s5)}^+$. In each of
(\ref{example1}) and (\ref{example2}), $R\rightarrow$const$>0$
as $\chi\rightarrow\infty$. There are no SS dust models
with $R\rightarrow 0$
as $\chi\rightarrow\infty$ [by (\ref{asymp1}) and since,
by (\ref{5sol}), $R\rightarrow 0$ requires $E\rightarrow 0$].
\[\]
{\bf ii. Open models with no origin}\\
By (\ref{asymp2}), to avoid a zero in $R$, a model with no origin must
either be composed entirely of (s6), or it must contain a section of (s5),
in order to allow (at least one) minimum in $R$.
Then the possible matchings are evident:
\[\left.\begin{array}{r}
{\cal S}\\
{\rm(s3)}^-{\rm(s5)}^-\\
{\rm(s4)}^-{\rm(s5)}^-\\
{\rm(s4)}^-{\rm(s3)}^-{\rm(s5)}^-\end{array}\right\}
{\cal S}\left\{\begin{array}{l}
{\cal S}\\
{\rm(s5)}^+{\rm(s3)}^+\\
{\rm(s5)}^+{\rm(s4)}^+\\
{\rm(s5)}^+{\rm(s3)}^+{\rm(s4)}^+\end{array}\right.\]
Examples and a detailed analysis of such models are provided in
\cite{hel2}. In these models, due to the presence of
collapsing solutions (s5),(s6), an origin does eventually form, but
gravitational collapse will violate the regularity conditions
in any case.
\[\]
{\bf iii. Closed models with two origins}\\
These models must contain a section of (s5), since there must
be (at least one) turning point in $R$.
The models cannot contain a section of
(s2), (s4) or (s2)(s4), since the section would either contain an origin
and match to another solution, or would match to other solutions on
both sides. Hence $E$ would vanish on both sides, and since $E>0$
throughout the domains of (s2) and (s4), $E^\prime$ could not have
the same sign throughout, contrary to (\ref{1234conds})
[with (\ref{mat4})]. There can be no (s1) section in the closed model,
since it does not match to any other solution. There can be no (s3)
region in the model either, since $R$ must increase in the direction
(s5)$\to$(s3). Hence if (s3) contains an origin, it cannot match
to (s5). Conversely, if (s3) does not contain an origin, it cannot
match to (s5) on both sides, leaving the model open.
This leaves just (s5) and (s6) to construct these models, and the
possibilities are:
\[{\cal O}{\rm(s5)}^+{\cal S}{\rm(s5)}^-{\cal O}\]
\[\]
{\bf iv. Closed models with no origin}\\
Consider an SS dust model which has $R>0$ in some range $0\leq r\leq d$
(and at some $t$).
This final possibility of composite models is obtained by identifying
(matching) the surfaces $r=0$ and $r=d$. Since $\Delta R=0$, the
model must be everywhere (s6) or else it must
contain a section of (s5)
[otherwise ${\rm sign}(R^\prime)$
is constant in $0\leq r\leq d$, which forces $R(0)\neq R(d)$].
No sections composed from the solutions (s1)-(s4) may be present, since
they would be forced to match to (s5) on both sides. This would force
$R^\prime$ to change sign in the section (since $R$ must increase
away from (s5) into these solutions) and this is not possible, by
(\ref{mat4}). Hence the models may only be
constructed from (s5) and (s6), with the possibilities:
\[ {\cal I}{\cal S}{\cal I} \]
where ${\cal I}$ denotes the surfaces which are identified (at which
the standard matching conditions must be satisfied, as we have
described). The spatial sections of these models
have the topology of a 3-torus. An example is provided by
\bea &&E=\left\{\begin{array}{ll}
ar^2-1      & {\rm for}~0<r<\frac{1}{4}d, \\
&\\
a(r-\frac{1}{2}d)^2-1 & {\rm for}~\frac{1}{4}d<r<\frac{3}{4}d, \\
&\\
a(r-d)^2-1 & {\rm for}~\frac{3}{4}d<r<d,
\end{array}\right.~~~
M=\left\{\begin{array}{ll}
b+cr^2 & {\rm for}~0<r<\frac{1}{4}d, \\
&\\
b+\frac{1}{8}cd^2-c(r-\frac{1}{2}d)^2 & {\rm for}~
\frac{1}{4}d<r<\frac{3}{4}d, \\
&\\
b+c(r-d)^2 & {\rm for}~\frac{3}{4}d<r<d,\end{array}\right.\nonumber\\
&& T=0,~~~a\left(2b+\frac{1}{4}cd^2\right)<\frac{4}{3}c,
\eea
where $a,..,d$ are positive constants. Note that a model of type
{\bf iv} cannot be constructed from the homogeneous
(Friedmann-Lema\^{\i}tre-Roberston-Walker) subclass of LTB
(since the elliptic homogeneous solution has only one point with $E=-1$,
at which $R$ is maximum).
\[\]
There are no further possible topologies or composite models.
Examples of models of types {\bf i}-{\bf iii} are given in
previous literature (see especially \cite{hel2}).

\section*{VIII. Conclusion}

In this paper, full regularity conditions have been derived and
discussed for SS dust spacetimes.
From section III, the solutions for the metric in section II
(which are all $C^\infty$ in $t$) may be joined only on comoving
surfaces. Hence the models are fully determined by choices of
the functions
$E(r)\geq-1$, $M(r)\geq 0$, $T(r)$, $A(r)$ and $B(r)$.
Existence of the Einstein tensor places a basic
restriction on these functions [equation (\ref{diffy})]:
$E$, $M$ and $T$ must be $C^2$, $A$ and $B$ must be
$C^1$, except at a finite number of points.
At these points, there may be discontinuities in any of
$E^\prime$, $M^\prime$, $T^\prime$, $A$ or $B$. However there are
additional, more subtle requirements of the functions. For example, at
points where $E\rightarrow -1$, equation (\ref{e=-1}) must hold.
These differentiability conditions ensure the good behaviour of
all the relevant physical quantities.

In the current context, the recollapse conjecture \cite{zel},
i.e. `all closed SS dust models must recollapse everywhere', follows
directly from the results of section VII. We simply note that
all the possible closed models contain only the solutions (s5) and
(s6), which both recollapse in a finite time. This is a simple
alternative proof to that of Burnett \cite{bur} (which involved
considering the lengths of timelike curves in these
spacetimes). Our proof slightly strengthens that of Bonnor
\cite{bon1}, in that no mathematical assumptions are required
(as were used in \cite{bon1})
other than those explicitly demanded by the regularity.
\[
\]
{\noindent{\bf Acknowledgements:} We thank Bill Bonnor and Charles Hellaby for helpful discussions. We thank Syksy Rasanen for reminding us about our original preprint. RM was supported by the South African Square Kilometre
Array Project, the STFC (UK) (grant no.
ST/H002774/1) and a Royal
Society (UK)/ National Research Foundation (SA) exchange grant.



\begin{thebibliography}{99}

\bibitem{bond} H. Bondi, Mon. Not. Roy. Astron. Soc. {\bf 107}, 410 (1947).

\bibitem{bon1} W. B. Bonnor, Class. Quant. Grav. {\bf 2}, 781 (1985).

\bibitem{hel1} C. Hellaby and K. Lake, Astrophys. J. {\bf 290}, 381 (1985). 
Erratum: {\em ibid.} {\bf 300}, 461 (1986).

\bibitem{Matravers:2000cu} D.~R.~Matravers and N.~P.~Humphreys,
   Gen.\ Rel.\ Grav.\  {\bf 33}, 531 (2001)
  [gr-qc/0009057].


\bibitem{Sakai:2008fi} 
  N.~Sakai and K.~T.~Inoue,
  Phys.\ Rev.\ D {\bf 78}, 063510 (2008)
  [arXiv:0805.3446].

\bibitem{Grenon:2011fs} C.~Grenon and K.~Lake, Phys. Rev. D{\bf 84}, 083506 (2011). 
[arXiv:1108.6320]

\bibitem{hel2} C. Hellaby, Class. Quant. Grav. {\bf 4}, 635 (1987).

\bibitem{mhms} R. Maartens, N. P. Humphreys, D. R. Matravers
and W. R. Stoeger, Class. Quant. Grav. {\bf 13}, 253 (1996) [arXiv:gr-qc/9511045]. Erratum: {\it ibid.}, 1689 (1996).

\bibitem{kam} E. Kamke, {\it Differentialgleichungen}
(B. G. Teubner, Stuttgart, 1977).

\bibitem{ell} G. F. R. Ellis, J. Math. Phys. {\bf 8}, 1171 (1967).

\bibitem{kra} D. Kramer, H. Stephani, M. A. H. MacCallum and E. Herlt,
{\it Exact Solutions of Einstein's Field Equations}
(Deutscher Verlag der Wissenschaften, Berlin, 1980).

\bibitem{isr} W. Israel, Nuovo Cimento {\bf 44B}, 1 (1966). Erratum: {\em ibid.}
{\bf 48B}, 463 (1967).

\bibitem{fay} F. Fayos, J. M. M. Senovilla and R. Torres,
Phys. Rev. D {\bf 54}, 4862 (1996). 

\bibitem{lak1} K. Lake, Phys. Rev. D {\bf 29}, 1861 (1984).


\bibitem{new} N. P. Humphreys, {\it PhD Thesis} (University of Portsmouth, 1998).

\bibitem{mtw} C. W. Misner, K. S. Thorne and J. A. Wheeler,
{\it Gravitation} (Freeman, San Fransisco, 1973).

\bibitem{pap} A. Papapetrou, Ann L'Institut Henri
Poincar\'{e} \underline{A} {\bf 25}, 207 (1978).
\bibitem{zel} Ya. B. Zel'dovich and L. P. Grishchuk, Mon. Not. Roy. Astron. Soc.
{\bf 207}, 23 (1984).

\bibitem{bur} G. A. Burnett, Phys. Rev. D {\bf 48}, 5688 (1993).



\end{thebibliography}
\end{document}